\documentclass{emulateapj}
\usepackage{epstopdf}
\usepackage{graphicx}
\usepackage{natbib}
\def\apjref#1;#2;#3;#4 {\par\pni\ #1,  #2, {\bf #3}, #4. \par}

\def\rsun{\ifmmode {\rm R}_{\mathord\odot}\else $R_{\mathord\odot}$\fi}
\def\msun{\ifmmode {\rm M}_{\mathord\odot}\else $M_{\mathord\odot}$\fi}
\def\lsun{\ifmmode {\rm L}_{\mathord\odot}\else $L_{\mathord\odot}$\fi}

\newcommand{\avg}[1]  {{\langle #1 \rangle}} 
\newcommand{\msd} {M_{\rm *d}}
\newcommand{\md} {M_{\rm d}}
\newcommand{\mdin} {\dot M_{\rm in}}
\newcommand{\csd} { c_{\rm s,d}}
\newcommand{\omkin} { \Omega_{\rm k, in}}

\newcommand{\Fig}[1]{Fig.~\ref{#1}}

\newcommand{\ltsim}{\protect\raisebox{-0.5ex}{$\:\stackrel{\textstyle <}{\sim}\:$}}
\newcommand{\gtsim}{\protect\raisebox{-0.5ex}{$\:\stackrel{\textstyle >}{\sim}\:$}}

\shorttitle{}
\shortauthors{Offner et al.}

\begin{document}

\title{The Formation of Low-Mass Binary Star Systems Via Turbulent Fragmentation}

\author{Stella S. R. Offner}
\affil{Harvard-Smithsonian Center for Astrophysics, 60 Garden Street,
  Cambridge, MA 02138, USA}
\email{soffner@cfa.harvard.edu }
\and
\author{Kaitlin M. Kratter}
\affil{Department of Astronomy and Astrophysics, 50 St. George Street, University of Toronto, Toronto, Ontario M5R 3H4, Canada}
\affil{ Institute for Theory and Computation, Harvard-Smithsonian Center for Astrophysics, 60 Garden Street,
  Cambridge, MA 02138, USA}
\and
\author{Christopher D. Matzner}
\affil{Department of Astronomy and Astrophysics, 50 St. George Street, University of Toronto, Toronto, Ontario M5R 3H4, Canada}
\and
\author{Mark R. Krumholz}
\affil{Department of Astronomy and Astrophysics, University of
  California, Santa Cruz, CA,  95064, USA}
\and
\author{Richard I. Klein}
\affil{Department of Astronomy, University of
  California, Berkeley, CA,  94720, USA}
\affil{Lawrence Livermore National Laboratory, Livermore, CA,  94550, USA}

\begin{abstract}

We characterize the infall rate onto 
protostellar systems forming in self-gravitating radiation-hydrodynamic
simulations. Using two dimensionless parameters to determine disks'
susceptability to gravitational fragmentation,
we  infer limits on protostellar
system multiplicity and the mechanism of binary formation.
We show that these parameters give robust predictions even in the
case of marginally resolved protostellar disks. We find that protostellar
systems with radiation feedback predominately form binaries via turbulent
fragmentation, not disk instability, and we predict turbulent
fragmentation is the dominant channel for binary formation for low-mass stars.
We clearly demonstrate that systems
forming in simulations including radiative feedback have fundamentally
different parameters than those in purely hydrodynamic simulations.

\end{abstract}
\keywords{stars: formation stars: protostellar disks, stars:low-mass}

\section{Introduction} \label{S:intro}
The study of stellar multiplicity is a centuries old pursuit \citep{Mitchell1767}, 
yet we still lack a comprehensive theory explaining the formation of the wide range of
observed binaries and multiple systems. In this paper, we investigate
the relative importance of two paths for binary formation using
simulations of a turbulent molecular cloud destined to form a small
cluster of low mass stars.

Although there are numerous proposed mechanisms for binary formation (see
for example \citealt{tohline02}), we focus here on two prominent
channels: the fragmentation of a turbulent core, and the
fragmentation of a gravitationally unstable disk. Briefly, the
turbulent core hypothesis \citep{goodwin04, fisher04} posits that turbulent
fluctuations within a bound core can produce multiple non-linear perturbations
in density, which exceed the local Jeans mass and collapse faster
than the background core. Multiple peaks within a given core result in
a bound binary or multiple stellar system. 

The second hypothesis, disk fragmentation \citep{ARS89,bonnell94},
suggests that disks subject to sufficiently strong gravitational
instability might fragment to form one or more
companions. Although previous studies suggested that this process was limited
to large mass ratio systems, recent work such as \cite{stamatellos09} and
\cite{kratter10} has shown that when disks continue to be fed at
their outer edges, the companions can grow substantially.

The latter process has shown promise in its application to high mass
stars \citep{kratter06,krumholz07a}, where rapid mass accretion pushes
disks towards instability. 
One of the goals of this paper is to investigate whether the conditions in low mass turbulent cores are ever sufficiently violent to induce
disk fragmentation.

Observations offer limited guidance as to which mechanism dominates,
although several lines of evidence indicate that turbulent
fragmentation must account for at least some of the binary
population. At wide separations binaries show a very broad
eccentricity distribution \citep{melo01} and random orientations
between the binary orbital plane and the spins of the component stars
\citep{hale94}. Similarly, disks around wide T Tauri star binaries show
significant misalignment between the planes of the disks and the
binary orbit \citep{jensen04, monin06, scholz10}. High
eccentricity and orbital misalignment are inconsistent with these
binaries having been created via the fragmentation of a single planar
disk
(unless they result from three-body interactions),
 but are expected from turbulent fragmentation. 
In contrast, binaries at small separations have reduced eccentricities and far less
misalignment between the orbital plane and the spin or disk
orientations of the individual components. However, this can plausibly
be explained via tidal circularization of initially misaligned systems
(e.g. \citealt{lubow00, bate00}), so these binaries may have
formed by either turbulent or disk fragmentation. 
Moreover, misalignment of the spin-orbital plane of systems born in disks may also be due to n-body interactions and ejections in a  multiple system.
Consequently,
observations are at present unable to answer the question of whether
disk fragmentation can be responsible for a significant fraction of
binaries. (See \citealt{howe09} for a further discussion of the
observational evidence on this point.)

Previous analytic work \citep{matzner05}, and numerical studies of
isolated disks have suggested that disks around low-mass stars will be
stable \citep{boley07, cai08}. This work has shown that appropriate
thermal treatment is essential, and in particular that heating by the
central star dominates at distances beyond 10's of AU.  
Recent numerical simulations of low-mass star formation have demonstrated
that small scale fragmentation leading to high multiplicity systems 
is indeed reduced when radiative feedback is included \citep{offner09,bate09}\footnote{\citet{bate09} includes radiative transfer on numerically resolved scales, but not radiation feedback from nuclear or accretion luminosity of stars, which Offner et al.~(2009) show is roughly an order of magnitude larger. In that respect Bate's simulations represent an intermediate case between non-radiative and radiative simulations.}. 

Despite these advances, resolving both the large-scale turbulence of a molecular cloud
forming an ensemble of stars and the details of disk structure remains
computationally challenging. In order to resolve the opacity limit for
fragmentation, SPH methods require mass resolution of $10^{-2} \msun$,
while grid based methods must achieve resolutions down to a few AU
\citep{goodwin07}.
Simulations often circumvent this difficulty by modeling individual
dense cores with simple initial conditions (e.g., \citealt{klessen00,
  goodwin04,walch09}). Such work is predicated on knowledge of
isolated cores properties
 and sacrifices the interaction of cores with their turbulent environment
\citep{offner08}.

In this paper we examine the simulations of Offner et al.~(2009b,
hereafter OKMK09), which follow the birth and evolution of protostars
in a turbulent molecular cloud, in light of the criterion for
protostellar disk fragmentation recently proposed by Kratter et al.~
(2010, hereafter KMKK10).   
%
We do this with the goal of drawing conclusions about aspects of stellar binary formation which are not resolved within OKMK09.   We proceed in two steps: first 
we test the KMKK10 stability criteria against non-isothermal, turbulent
simulations using the high-resolution runs from OKMK09; 
 then we
apply these criteria in order to predict the outcome of under-resolved
disk accretion in the low-resolution OKMK09 simulations.  
Ultimately, we can assess
the relative importance of turbulence and disk instability in the creation of 
low mass stellar binaries.

In the upcoming sections we define our key dimensionless parameters, outline our testing scheme, describe our numerical methodology, and ultimately draw conclusions about the nature of low-mass star formation.

\section{Characterizing Accretion and Disk Fragmentation}  \label{dimen_pars}
\subsection{Dimensionless Parameters}
Classically, disks are thought to fragment and become unstable when
self-gravity becomes sufficiently strong as measured by 
a low value of 
Toomre's $Q$ parameter:
\begin{equation}
Q = \frac{\csd \Omega_{\rm d}}{\pi G \Sigma_{\rm d }},
\end{equation}
where $\Omega_{\rm d}$ is the disk angular velocity, $\Sigma_{\rm d}$ is
the disk surface density, and $\csd$ is the disk sound speed \citep{toomre64}. More
recent work has shown that disk gas must also
be able to cool efficiently once the process of collapse begins in
order to fragment \citep{gammie01}. The latter criterion is a strong constraint 
in the inner parts of disks where viscous heating dominates
but is typically satisfied in the outer radii of the irradiated disks
that we study here \citep{kratter08,kratter10b}.

Although $Q$ remains a good predictor of disk fragmentation, it is of
more limited use for evaluating disk stability in both
large-scale star formation simulations and observations because it requires precise
knowledge of local disk properties, which are difficult to model and
measure.

Young protostellar disks are typically driven unstable only when they
receive mass faster than they can process it down onto the central
star \citep{matzner05}. Thus it is useful to parametrize disk
fragmentation as a function of the infall rate from large scales.
 KMKK10 demonstrated that one can predict disk fragmentation under
idealized conditions using two dimensionless numbers normalized to the
infall rate. One, the thermal parameter $\xi$, compares the disk sound speed to the mass accretion rate:
\begin{equation} \label{eq-defxi}
\xi = {{\mdin G}\over{\csd^3}},
\end{equation}
where $\mdin$ is the infall mass accretion rate.  
This relates to the  \cite{SS1973} $\alpha$ parameter through $\xi = 3
\alpha/Q$.
Increasing $\xi$ at fixed $\alpha$ tends to cause $Q$ to decrease until self-gravity becomes strong ($Q\sim 1$), at which point gravitationally induced spiral modes lead to an increase in the effective value of $\alpha$.   Beyond critical values of $\xi$ and $\alpha$, however, the disk will fragment; practically this occurs when $\xi \simeq 2$ and $\alpha \simeq 2/3$.    For instance, a strictly isothermal simulation fed by a slowly rotating \citet{shu77} solution, in which $\xi=0.975$, should show strong spiral arms but not fragment, whereas one fed by the \citet{foster93} collapse solution, in which $\xi \gg 1$ at early times, should fragment. 

A second parameter, $\Gamma$, measures rotation by comparing the orbital period of the infalling gas to the accretion time scale:
\begin{equation}
\Gamma = {{\mdin}\over{\msd\omkin}} = {{\mdin \avg{j}_{\rm in}^3}\over{G^2 \msd^3}}, 
\end{equation}
where $\msd$ is the total mass in the star-disk system, $\omkin$ is
the Keplerian angular velocity at the circularization radius of the infall, and $\avg{j}_{\rm in}$ is the 
average
specific angular
momentum.  A large value of $\Gamma$, e.g., $10^{-1}$, implies that the system mass
changes significantly in only a few disk orbits, whereas small values,
$\Gamma \sim 10^{-4}-10^{-3}$, describe disks with a mass doubling time
of hundreds to thousands of orbits.   The magnitude of $\Gamma$ affects the disk's aspect ratio ($H/R_d \simeq (\Gamma/\xi)^{1/3}$), and therefore the winding of spiral arms, and the mass of forming fragments.

Conducting idealized collapse experiments with ORION, a 3-D Adaptive Mesh Refinement (AMR) gravito-radiation-hydrodynamics code, KMKK10 probed the fragmentation threshold and final outcome (single, binary, or multiple) for isothermal disks in which $\xi$ and $\Gamma$ are constant but numerical resolution improves over time.   They concluded that any disk with $\xi > 2-3$ will fragment,  for values of $\Gamma$ typical of low mass star formation($10^{-3} \ltsim \Gamma  \ltsim 10^{-2}$),  and that increasing $\Gamma$ has a weak stabilizing effect.   

KMKK10 argued that $\xi$ and $\Gamma$ capture the essential aspects of disk thermodynamics in the context of steady accretion, and that other parameters, such as the cooling time compared to $\omkin^{-1}$, influence fragmentation only through their effect on $\xi$. This assertion is not tested within their isothermal simulations, but in \S4 we confirm that it is consistent with the behavior of our highest resolution (non-isothermal) runs.  We then treat $\xi$ and $\Gamma$ as robust predictors of unresolved disk instability and fragmentation in our lower resolution simulations. 
%

\section{Methods} \label{methods}

\subsection{Numerical Methodology} \label{comp}
  
In this section, we give a brief overview of the simulations completed
in OKMK09 that are the basis for this work. The calculations are
performed with the ORION AMR code. 
The
simulation boxes have initially uniform density and  periodic boundary
conditions. Energy is injected in the form of small velocity
perturbations with wave numbers in the range $1\le k\le 2$ for three
crossing times until a turbulent steady state is achieved. 
The calculation we refer to as RT includes radiative transfer in the
Flux-Limited Diffusion approximation, while the second calculation,
NRT,  uses a barotropic equation of state that has no mechanism for
the transport of radation. 
Since radiative cooling is
very efficient during the initial driving phase, all of the gas
remains close to 10 K in the RT run, and both calculations begin with
a similar temperature.

After three crossing times, self-gravity is turned on and this point
is considered $t=0$. The initial mean density is $4.46\times10^{-20}$
g cm$^{-3}$, the 3D Mach number is 6.6, and the total mass is 185
$\msun$.  Energy continues to be injected at a constant rate to offset
the natural turbulent decay throughout the run
(e.g. \citealt{stone98}). 
Stars are inserted as point particles once the Jeans criterion is
exceeded on the maximum level \citep{krumholz04}. In the RT
calculation, the stars are endowed with a subgrid stellar evolution
model, which includes accretion luminosity down to the stellar
surface, Kelvin-Helmholz contraction, and nuclear burning (see OKMK09
for a detailed description). The calculations evolve with gravity for
one cloud freefall time or 0.315 Myr.

The RT and NRT runs have a minimum AMR cell size of 32 AU such  that
protostellar disks are only marginally resolved with $\sim 10$
cells. Although the
simulations refine based upon the Jeans length, thus
preventing artificial fragmentation \citep{truelove97}, they do not
resolve the 
 disk scale height 
with more than a few cells
\citep{nelson06}. Nonetheless,
OKMK09 demonstrated that the protostellar accretion rates are
converged to within a factor of two by running a resolution study of
the first forming protostar in each with an extra three levels of AMR
refinement and minimum cells spacing of $\sim$ 4 AU. We use the high resolution studies RTC and NRTC to assess convergence. 

\subsection{Data Analysis} \label{SS:data_analysis}

Our parameterization in terms of $\xi$ and $\Gamma$ requires that we distinguish disk matter from the infall, so that we can evaluate the disk sound speed $\csd$ and the angular momentum scale $\avg{j}_{\rm in}$ in addition to the accretion rate $\mdin$. 
The first step is to define the disk-accretion boundary in a robust way within our simulations. 
Imposing a density
cutoff is not sufficient as the protostellar cores have a range of
conditions, which produce diverse disk properties. Automatic
identification is complicated by disk flaring, close companions and
turbulent filaments of gas feeding the disks. For the RT and NRT
calculations we adopt a relatively low density threshold of $10^{-16}$
g cm$^{-3}$. We estimate the total angular momentum vector of the gas,
and rotate the coordinate frame so that 
the net angular momentum vector is parallel to the z axis.
Finally, we restrict the vertical disk height in the z
direction to $\pm 5$ cells from the disk midplane. The combination of
a gas density cut with these geometric constraints captures the
 flaring and warping of the disk, while excluding gas flowing into the
disk. These criteria are not sufficient to remove abutting disks of
very nearby companions from the sample, but these events are few and
are easily identified from changes in the estimate of the disk
radius. 
In the case of a multiple system with a single disk 
we perform the analysis in the
reference frame of the primary.  

Once the disk material is identified, we estimate the disk mass, radius, mean
temperature, accretion rate and angular momentum. The disk radius,
$r_{\rm d}$, is defined by the distance between the farthest disk
cell and the protostellar location. The mean sound speed, $\csd$, is
given by $\sqrt{k_{\rm B} T_{\rm d}/\mu_{\rm p}}$, where $\mu_{\rm p}
= 2.33 m_{\rm H}$ is the mean particle mass and $T_{\rm d}$ is the
mean mass-weighted temperature averaged over the disk. We find that
constructing a volume weighted mean changes the resulting temperature
by at most 15\%. The accretion rate, $\mdin$,
is calculated by taking the difference between the total star-disk system
mass, $\msd$, 
from one time step to the next.
 In rare cases,  $\mdin$ may be negative if
the disk is perturbed by a nearby companion or passes through a
shock.

We find that varying the density threshold by factors of two has a
10-15\% effect on the disk mass and accretion rate but  may translate
into a 
50\% difference in $\Gamma$ and $\xi$, since these
parameters are implicitly sensitive to the disk radius and the amount
of included or excluded high angular momentum material near the disk
boundary. 
The RTC and NRTC runs, which have better
resolved disk structure and sharper disk edges, serve as limits on the
sensitivity to specific disk properties.
We discuss in more detail in \S\ref{corevals} how the parameters change
when defined on larger scales.

\section{Results} \label{results}
\subsection{Final Protostellar System Outcomes }

\begin{figure*}
\epsscale{1.10}
\plotone{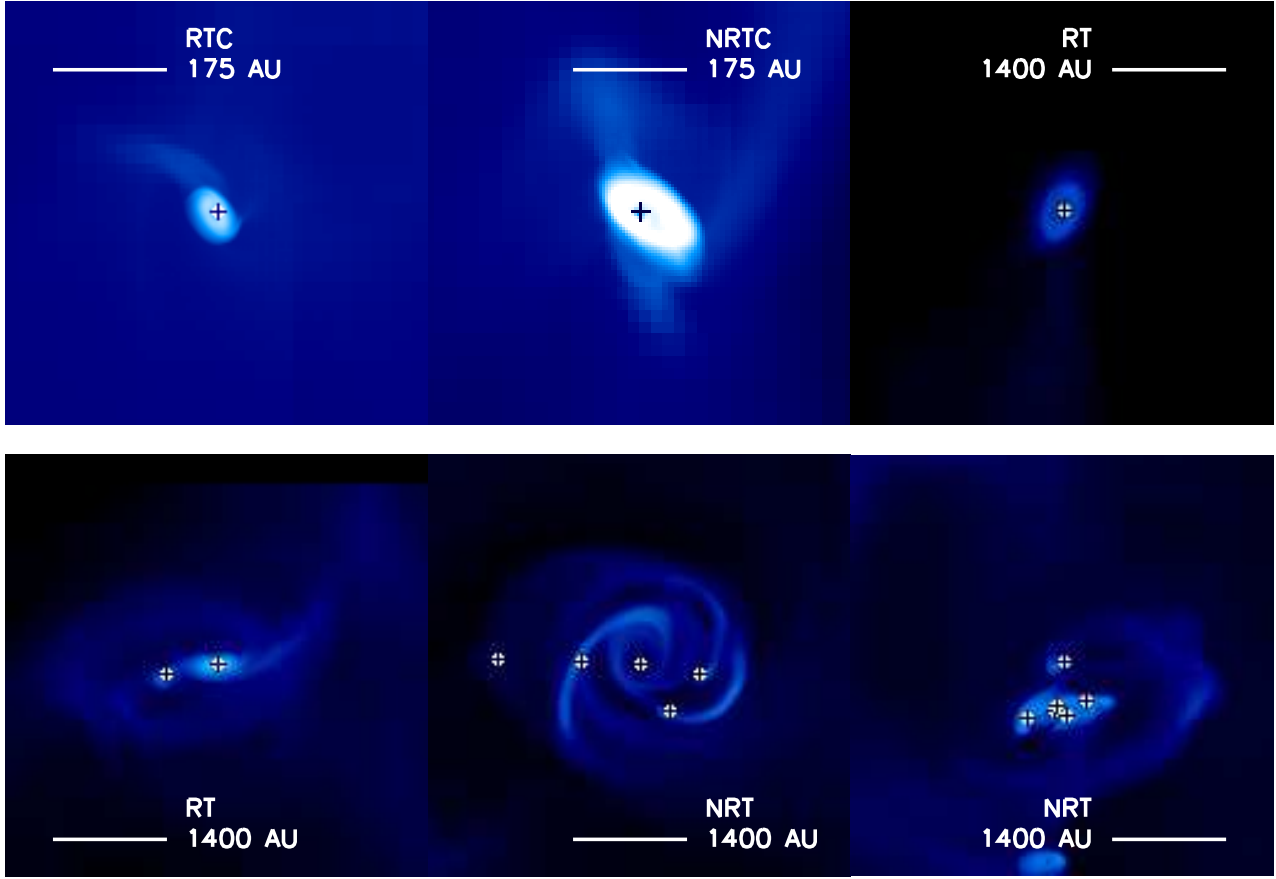}
\caption{ \label{diskimg}
Log gas column density for six different protostellar systems, where
the star positions are marked by crosses. The column
density scale runs from 0.1 gcm$^{-2}$ to 100 gcm$^{-2}$.  The scale
of the image and run are indicated.
}
\end{figure*}

Figure \ref{diskimg} shows examples of several systems formed in each of the runs (RT, NRT, RTC, and NRTC).
The systems can be divided into three classes: isolated stars, binaries and multiples formed via the fragmentation of a turbulent core, and binaries and multiples formed via the fragmentation of a disk.     Because disk radii are of order a few hundred AU, whereas turbulent fragmentation of a core can occur on scales up to the Bonnor-Ebert radius ($0.05$\,pc $=10^4$\,AU for $n_H=10^5$\,cm$^{-3}$, $T=10$\,K), we use 500 AU as a crude dividing line. 
%
 
\begin{figure}
\epsscale{1.10}
\plotone{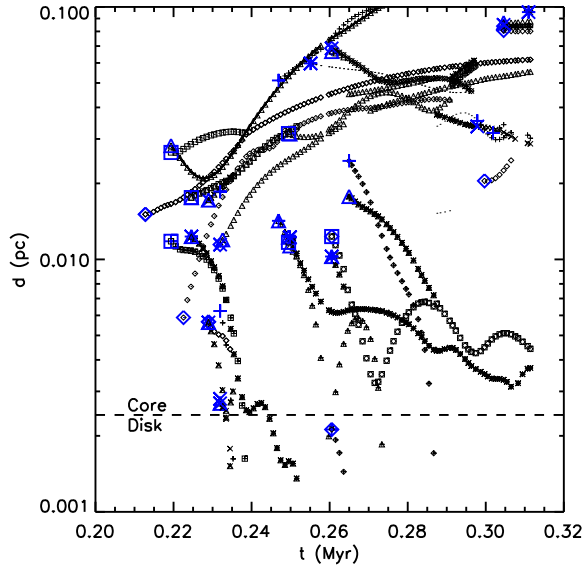}
\caption{ \label{dvst}
Pair separation as a function of time in 1 kyr bins for pairs in the RT
simulation, 
where each symbol indicates a different pair.
The dashed line at 500 AU indicates a rough boundary between
the disk and core scale,where
the mean disk size in the RT simulation is approximately 500 AU.
 The large majority of pairs have separations above 0.1 pc
and are not shown here. The large (blue) symbols indicate the first
time bin.
}
\end{figure}

 In Figure \ref{dvst}, we show the separation of all possible pairs of stars, bound or unbound, as a function of time in the simulations including radiative feedback (RT).  All but one binary system form outside
%
of 500 AU, but within a typical core radius. 
Upon inspection, the 
exception results not from disk fragmentation but rather from
fragmentation of a cold filament or 
stream feeding a more massive protostar. 
We conclude that
fragmentation leading to binaries in the RT run is consistent with core, not disk, fragmentation. 

Note that the pairs that appear to diverge in Figure \ref{dvst} are unaffiliated stars, not spreading binaries. Pairs with
separations greater than 0.1 pc are excluded since they 
clearly form in different cores and cannot result from turbulent core fragmentation. 

The
fragmentation history in the calculations without feedback is
quite different.  
Turbulent fragmentation still occurs on core scales, but now 
protostellar disks also fragment; see Figure \ref{diskimg}. 
As discussed in OKMK09, 
and as predicted by \citet{matzner05},
this additional  fragmentation would have been suppressed by realistic heating of
the disk via radiative feedback.   In run NRT, which lacks such heating,
disk fragmentation leads to several large multiple systems and the dynamical ejection of a few of the smaller
companions.
Considering that radiative feedback is present in low-mass star formation, we
 conclude that turbulent core fragmentation is likely responsible for the low mass stellar binary population.

%

\subsection{Simulation Thermal and Rotational Parameters: $\xi$ and $\Gamma$}

\begin{figure}
\epsscale{1.10}
\plotone{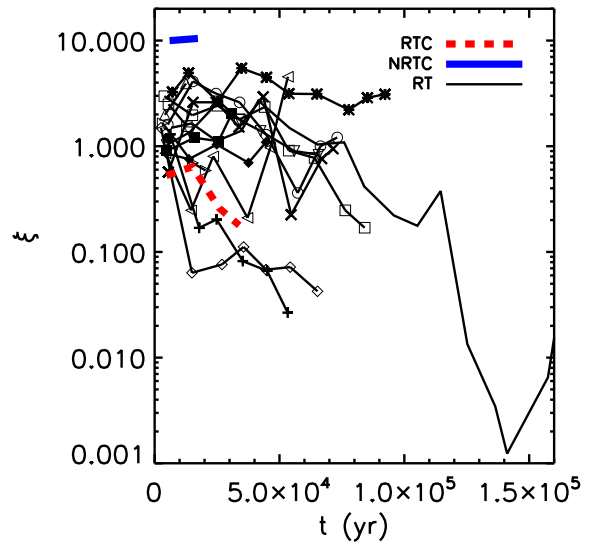}
\caption{ \label{xvst}
The thermal parameter, $\xi$, as a function of time for the protostellar
disks in each simulation. The high-resolution RTC (dashed, red) and NRTC
(solid, blue) runs are shown in bold. The solid line without a symbol
corresponds to the same first forming star that is depicted by the RTC
and NRTC runs. The data is averaged over $10^4$ yr bins.
}

\end{figure}

We find that the thermal parameter, $\xi$, as illustrated in Figure
\ref{xvst} is either 
constant or decreasing with time for each system. As shown, most
protostars begin with mean values of  $\sim$ 1-3, skirting the regime where fragmentation is possible. 

The decline in $\xi$ over time primarily reflects that the accretion rates decrease as the disk temperatures do not increase significantly. This is consistent with
observations that the mean accretion rate falls as protostars
transistion from the Class 0 to Class II stages \citep{andre94, evans09}. 
By the time protostars reach the Class II stage, their luminosities
are dominated by stellar rather than accretion luminosity  \citep{white04}.

The $\xi$ value from the NRTC run is shown 
in Figure \ref{xvst}
for reference. It is apparent that without 
heating from
radiation feedback 
$\xi$ reaches a 
value that is 
 a factor of 10 and 20 times
larger than its RT and RTC counterparts, respectively. The RTC $\xi$
begins within a factor of 2 of $\xi$ for the corresponding RT
protostar. However, the more resolved disk is smaller and has a higher
mean temperature so that the values differ by a factor of nearly
10 at 40 kyr. Because 
$\xi\propto T^{3/2}$, 
any temperature difference is magnified.

The RT disks cluster around mean temperatures of 30-40 K.
Since the protostellar luminosity is comprised of both a stellar and
accretion component, heating continues and the disks remain fairly
warm even in those cases that the accretion rate, $\mdin$, diminishes
to $\simeq 10^{-7} \msun$ yr$^{-1}$. 
In contrast, the disk in the NRTC run, which has a barotropic EOS,
remains close to an average temperature of 10 K,
 a factor of $\sim$ 4 smaller than in the cases with radiative feedback.

\begin{figure}
\epsscale{1.10}
\plotone{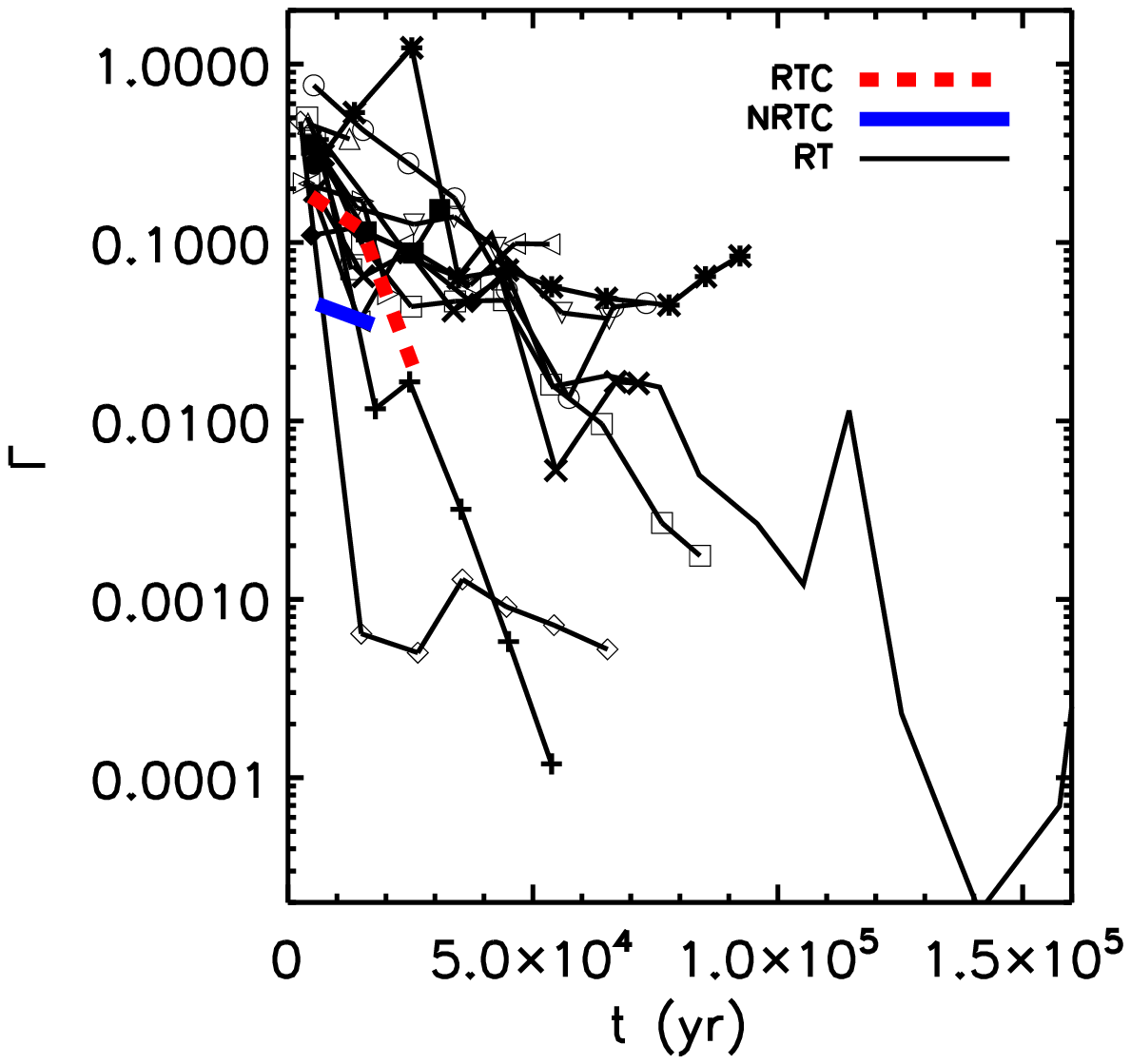}
\caption{ \label{gamvst}
Rotational parameter, $\Gamma$, as a function of time for the protostellar disks in each simulation. The lines, symbols, and data bins are the same as in Figure \ref{xvst}. 
}
\end{figure}

Figure \ref{gamvst} illustrates that the 
rotation parameter
 $\Gamma$, 
drops rapidly in time. 
For the first $\sim 10^4$ years,
the
 decline arises from increasing disk masses,
as high angular momentum material settles onto the disk
faster than the disk can drain material onto the star.

At later times, decreasing $\Gamma$ 
results from declining accretion rates as the core gas is
 depleted. 
This trend can be seen in Figure \ref{gamvst} for a number of
protostars that have evolved for more than $5\times 10^4$ years.
The NRT and RT $\Gamma$ values are more similar and fall
 within the same range of parameter space, because $\Gamma$ depends
 primarily on the turbulent initial conditions on large scales, which
 are mostly unaffected by radiative feedback.
Both the $\Gamma$ and $\xi$ curves exhibit similar shape variation as a
function of time due to their linear dependence on the
accretion rate.

Figure \ref{gamvsxi} shows 
$\Gamma$ as a function of $\xi$ at different times for the
protostellar disks in each simulation.
The values fall in a fairly narrow strip of parameter space. Although
 fluctuations in the variables obscure the trend somewhat, protostars
 move from the top right of the plot to the bottom left. 
For comparison, we 
plot the $\Gamma-\xi$ tracks
 for collapsing Bonnor-Ebert spheres initially are overdense by 10\%
 \citep{foster93}. 
(We assume their disks are heated
by a factor of 3.5 above the core temperature,
consistent with the average of the RT runs.)
The $\Gamma$ values for the Bonnor-Ebert spheres are calculated assuming solid-body
rotation with rotational parameters $\beta = 0.02$ and $\beta =0.08$, where  $\beta = E_{\rm rot}/ E_{\rm grav}  $. 
Both cases lie within the RT data and 
mimic
 the evolution towards lower
$\xi$ and $\Gamma$ with time.

\subsection{
Test of Fragmentation Criteria} 
\label{tests}
\begin{figure}
\epsscale{1.10}
\plotone{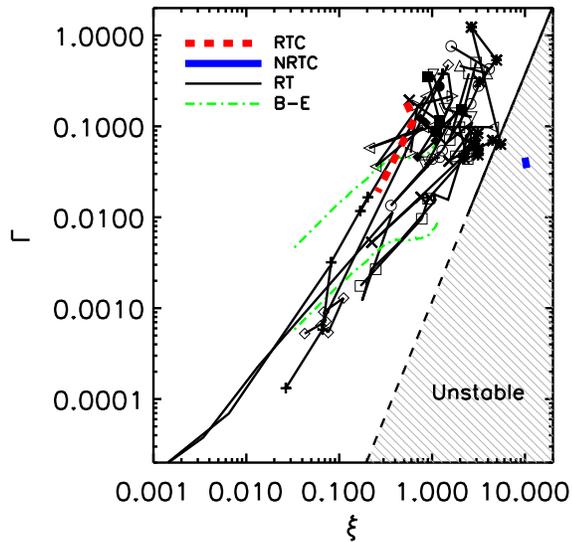}
\caption{ \label{gamvsxi}
$\Gamma$ as a function of $\xi$ for the protostellar disks in each
  simulation.  The solid line
  is the boundary between fragmenting and stable disks found by
  KMKK10. The dashed line indicates where we have extended it to
  lower $\Gamma$ values than explored by KMKK10.  Bonnor-Ebert (B-E) spheres
  for $\beta = 0.08$ (top) and  $\beta = 0.02$ (bottom) are given by
  the (green) dot-dashed lines.
The lines, symbols, and data bins are the same as in Figure
  \ref{xvst}.
}
\end{figure}

With the $\xi$ and $\Gamma$ values in hand we can compare the system outcomes to those predicted by the idealized models of KMKK10. We take advantage of both the NRT and RT runs to explore disk outcomes across the fragmentation boundary in $\xi-\Gamma$ space. 
KMKK10 predict
systems that fall to the left of the solid line in Figure
 \ref{gamvsxi} should be stable to disk fragmentation, while systems to the right should fragment. This is precisely what we find; disks to the right of the line, which are primarily NRT runs, undergo disk fragmentation, while those to the left, primarily RT runs, do not.  Note that low resolution
often prevents 
the fragments from surviving to form binaries due to the sink particle algorithm (see \S\ref{sec-res}); a similar effect was observed in resolution studies in KMKK10. 
We find that in a few cases, RT disks briefly cross the fragmentation line as a result of short duration accretion variability.

\begin{figure}
\epsscale{1.10}
\plotone{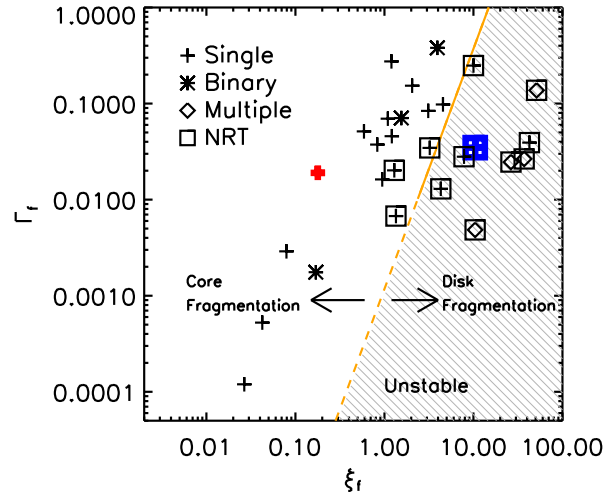}
\caption{ \label{fingamvsxi}
The final values of $\Gamma$ and $\xi$ at 1 $t_{\rm ff}$ for the
protostellar disks in each simulation. The high-resolution RT (red)
and NRT (blue) runs are shown in bold.  The hatched area indicates
$\xi$ and $\Gamma$ values prone to disk instability as shown in Figure \ref{gamvsxi}.}
\end{figure}

Figure \ref{fingamvsxi} shows the final value of $\Gamma$, $\xi$, and
multiplicity for each protostellar system. Here we define a multiple
system as one with bound neighbors within 2000 AU.
The RT cases all fall to the left of the fragmentation line, and all
binaries 
in this region are formed via core fragmentation. In contrast, the NRT systems are nearly all to the right of the
fragmentation line, with the multiple system having the highest $\xi$
values. Although there are single NRT systems on both sides of the line, we find that those in the fragmenting region have previously or are still experiencing fragmentation and particle merging.
%
 The formation of primarily single systems in the RT run is consistent with the observed multiplicity fraction of low mass stars \citep{lada06}.

\subsection{ Toomre's Q and Disk to Star Mass Ratios}
\begin{figure}
\epsscale{1.10}
\plotone{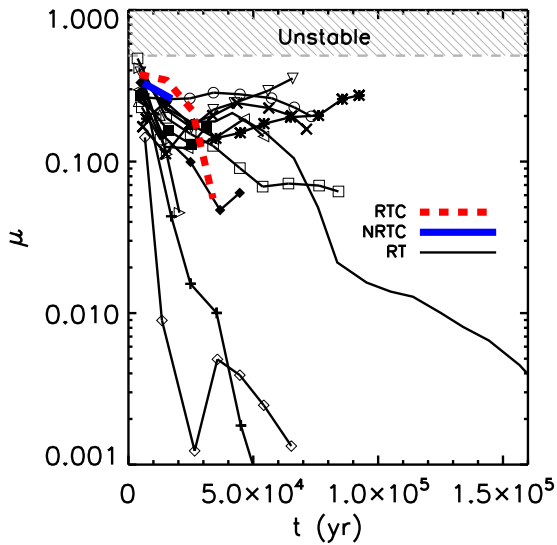}
\caption{ \label{muvst}
The disk to system mass ratio $\mu$ as a function of time for the
protostellar disks in each simulation. The lines, symbols, and data
bins are the same as in Figure \ref{xvst}. The hatched area indicates
values of $\mu$ prone to disk instablity.}
\end{figure}

Although we rely primarily on $\xi$ and $\Gamma$ to predict disk
instability we also examine  both the disk to star mass ratio
and Toomre's $Q$ parameter as proxies for disk instability.
Following \cite{kratter08}, we define the ratio of the disk mass to total system mass as 
\begin{equation}
\mu = {{\md}\over{\msd}}.
\end{equation}
When the disk and star
are of comparable mass, or $\mu \simeq 0.5$, we
expect that the disk will be unstable to gravitational fragmentation 
(e.g., \citealt{shu1990,kratter08}). Figure \ref{muvst} shows $\mu$ as a function of time.

\begin{figure}
\epsscale{1.10}
\plotone{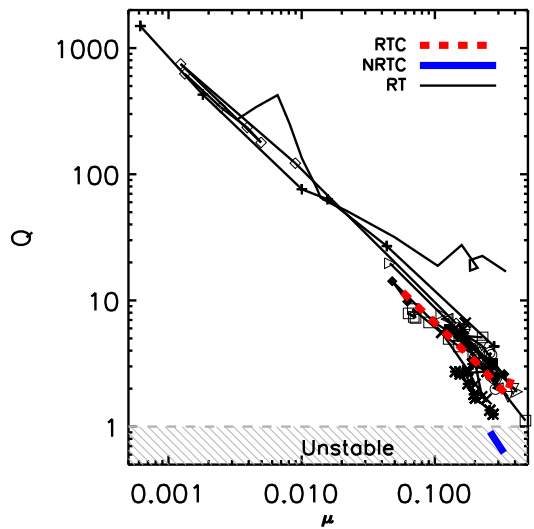}
\caption{ \label{qvsmu}
Toomre $Q$ parameter as a function of the ratio of disk to system
mass, $\mu$, for the protostellar disks in each simulation. The
high-resolution RTC (dashed) and NRTC (solid) runs are shown in bold.  
The hatched area indicates
values of $\mu$ and $Q$ prone to disk instablity.
}
\end{figure}

To calculate a disk-averaged value for $Q$ we use the density weighted
temperatures to calculate the disk sound speed. We estimate the
disk-averaged surface density as 
$\Sigma_{\rm d} = \md / (2 \pi r_{\rm d}^2)$, 
where the coefficient $2$ 
assumes 
a $1/r$ disk column density profile. Because very massive disks ($\mu \sim 0.5$) are
especially prone to fragmentation, we show the trajectory of disks in
$Q-\mu$ space in Figure \ref{qvsmu} for the disks in each of the
simulations. 
Thin disks with $Q \lesssim 1$ are unstable to fragmentation. Note
that thicker disks fragment at lower values of $Q$ (of order 0.7),
while non-axisymmetric gravitational instabilities can set in at
$Q\sim 2$ \citep{GLB1965,sellwood84}.

 At early times, just after the onset of collapse, the protostellar mass is
small compared to the disk. However, this phase is brief,
approximately a few $10^3$ years, and these structures may in fact
be underesolved flattened envelopes 
rather than rotationally supported disks.
After this early phase, the systems settle into a state where the disk is 
approximately one quarter of the system mass.

 As the core mass is accreted and the protostar grows, $\mu$
declines. Both $\Gamma$ and $\xi$ are positively correlated with the
behavior of $\mu$ since a declining disk-system ratio signals a
declining infall rate.

As suggested by Figure \ref{qvsmu}, disks in the RT and RTC
simulations have $Q > 1$ without exception and thus lie in
the stable part of the $Q-\mu$ parameter space.  
In contrast, the NRTC disk approaches $Q\simeq 1$ from below, suggesting that it is
extremely unstable at early times. As expected, early fragment formation occurs in
the NRTC disk during the first several kyr of the simulation 
when $Q \simeq 0.7$. 
As shown in Figure \ref{muvst}, $\mu$ generally decreases with time. In
Figure \ref{qvsmu}, 
declining $\mu$ corresponds to larger $Q$ values and
increased disk stability.

\subsection{Semi-Analytic Comparison}
\label{sec-analytic} 

 Although we measure disk properties directly
in these simulations, certain properties remain unresolved at lower resolution and
are likely affected by artifacts such as numerical diffusion. 
In addition,  the previous metrics represent global averages, which may disguise smaller scale instability.
%

To examine the importance of resolution and global averaging, 
we make a second estimate of disk stability
using simple analytic models to predict 
radius-dependent disk
properties, while still relying on the well resolved infall rates and the luminosities
determined in the simulations.

\begin{figure}
\epsscale{1.2}
\plotone{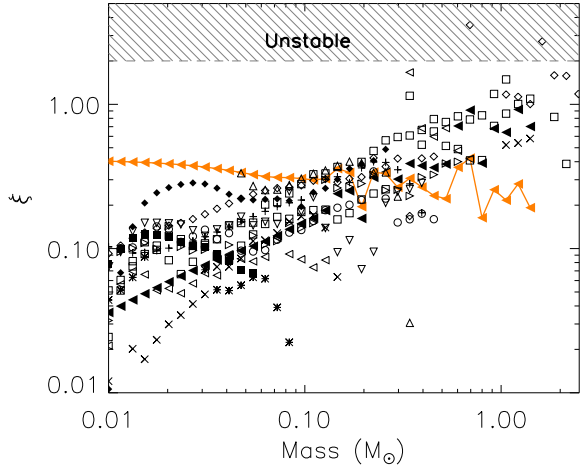}
\caption{ The value of $\xi_{\rm d,ir}$ calculated for the estimated
  irradiated disk temperature,  $T_{\rm d,ir}$, at a mass dependent
  radius of $r_d = 200 ({M_*}/{\msun})$ AU for each particle
  (black).  For comparison we also show the $\xi$ trajectory for one
  of the particles at a fixed radius of $50$ AU (triangles connected with solid, orange line). 
The hatched region indicates approximate values of $\xi$ at which systems will become unstable for values of $10^{-3} <\Gamma < 5 \times 10^{-2}$.}
\label{fig-all_mass}
\end{figure}

Disks are driven unstable when they are fed material more rapidly than they can process it at a given temperature, i.e. when $\xi \gtsim 2$, as discussed in \S~\ref{dimen_pars}. 
%
To evaluate $\xi$ 
we estimate disk temperatures at characteristic radii using the disk
irradiation models of \cite{matzner05}. They find that embedded disks
absorb a large fraction of the emitted starlight because the infall
envelope is optically thick to visible wavelengths which are caught and
re-emitted towards the disk in the infrared. Using a ray tracing
calculation they find that the flux reaching the disk surface is
approximately:
\begin{eqnarray}\label{ml05flux}
F_{d} = \frac{f_{*d} L_*}{4 \pi r_d^2} \rm{~where} \\
f_{*d} = 0.11\epsilon^{-.35},
\end{eqnarray}
and $\epsilon$ is the accretion efficiency from the core onto the
star-disk system.  We adopt $\epsilon
= 1/3$ here (see discussion in \S\ref{simp}). 
The temperature of a disk in which irradiation dominates over viscous heating, and which is optically thick at all relevant wavelengths, will be
\begin{equation}
T_{\rm d,ir} = \left(\frac{f_* L_*}{4\pi r_d^2 \sigma}\right)^{1/4}.
\end{equation}
This estimate is realistic, since disks near their fragmentation threshold are indeed likely to be optically thick 
\citep[][ discussion below their eq. 37 ]{matzner05}, and because viscous heating is minimal beyond a few tens of AU
\citep{kratter10b}.
In our evaluation we use $L_*$ as calculated in our simulation; although this is somewhat affected by unresolved dynamics through its dependence on the accretion rate, we believe the error to be small. 
%
And, since our goal is to evaluate stability in a way that is independent of poorly resolved disk radii, we consider a definite radius-mass relation,
$r_d = 200 ({M_*}/{ \msun})$ AU. This scaling follows from the
assumption that the disk radius scales with the core radius; 
if cores share a common turbulent Mach number, then their maximum disk
size is proportional to $R_{\rm core}$, although there may be large
fluctuations around this trend.  When cores are pressure confined and supported by both thermal pressure and
subsonic turbulence, then core radius, and thus disk radius, scales linearly with mass at fixed temperature. Note that the form of this relation is relatively unimportant as disk stability is determined by the maximum size to which the disk grows
\citep{matzner05}.
We adjust the normalization of this scaling to match our high resolution runs. 
%
 In \Fig{fig-all_mass} we plot $\xi$ for each star in the RT run as a function of the star's current mass. 
For comparison we also show $\xi$ calculated at a constant radius of
 $50$ AU for one of the stars.  The overall variability of $\xi$ is due to
 variable accretion rates and thus stellar luminosities, while the
 increase with mass is due primarily to a decrease in
 temperature as the characteristic radius increases. 
 At low masses, the analytic radii are significantly smaller than those in the simulations, which are somewhat enlarged due to numerical diffusion; in reality disks may be larger and less coherent at early times than analytic models predict. 
We find that the radius normalization must be increased to at least
 600 AU, a very large disk size for these stellar masses, before a significant number of $\xi-M$ values move into the unstable
 regime.
 This result is consistent with the global trends of
 $\xi$ and $\Gamma$ derived in the previous sections.

Stabilization out to such large radii is due to strong external
radiation at early times. Otherwise disks become unstable outside of a much
smaller radius,
within which cooling times are too long and viscous heating alone suffices to suppress fragmentation 
\citep{rafikov05,matzner05,clarke09}


\section{Discussion}\label{discussion}

\subsection{Sensitivity to Parameter Definitions}\label{corevals}

We use the parameters $\xi$ and $\Gamma$ as proxies for gravitational fragmentation, but as we discussed in \S~\ref{SS:data_analysis}, our evaluation of these parameters depends somewhat on our ability to discriminate disk from infall.  
%
%
To gauge what uncertainty this may cause, we 
we reevaluate $\Gamma$ and $\xi$ using $T_d$, $\avg{j}_{\rm in}$, and $\mdin$
averaged on (or within) a sphere of radius 
1000 AU centered on the protostar, well outside the actual disk but close enough that the enclosed mass is dominated by disk material.
%

\begin{figure}
\epsscale{1.10}
\plotone{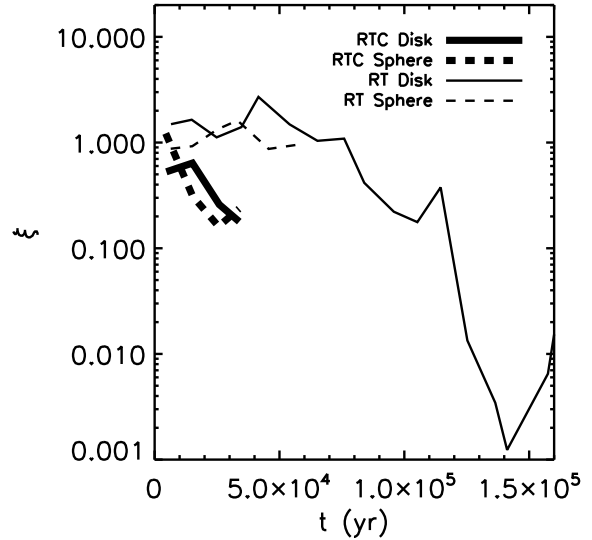}
\caption{ \label{convxi}
The thermal parameter, $\xi$, versus time for the first forming
protostar in the RT and RTC (bold) calculations using the fiducial
disk definition (solid) and a 1000 AU sphere (dashed). 
}
\end{figure}

Figure \ref{convxi} shows $\xi$ for the first forming protostar in the
RT and RTC simulations with the parameters estimated from the fiducial
disk definition and from a sphere with 1000 AU radius. The $\xi$
values for the two different geometries follow a similar progression
and they are generally within a factor of two. This confirms that the
accretion rate on small scales is driven by the larger scale
characteristics of the core and is not particularly sensitive to the
details of the disk geometry 
or analysis. The difference between the RT and RTC
case is preserved using the spherical geometry and is determined
mainly by the different mean temperatures in the two cases (see
\S3).

 We find similar agreement between
the two methods of evaluation when this analysis is repeated for $\Gamma$: the RT
rotational parameter in the disk and sphere geometries is within a 
factor of $\sim$2. The two RTC $\Gamma$ values follow
a similar trend and lie within a factor of 3.

\subsection{Simplifying Assumptions}\label{simp}

The caveats of the numerical methods are discussed in detail by
OKMK09.  We revisit two main caveats here in the context of these results.

Magnetic fields are indisputably important in star formation. Certainly they play a role in launching outflows, which we discuss below, and also serve as a source of pressure that may resist and slow the influence of gravity. 
Methods using 3D ideal magnetohydrodynamics (MHD) find that the
inclusion of magnetic fields suppresses disk fragmentation in the
parameter space applicable to low-mass star formation \citep{price07,
  hennebelle08}, a result that is complementary to the role we find
for radiative feedback. 
These simulations suggest that even the formation of a disk may be
suppressed if the field is strong and the angular momentum vector is
aligned with the field.
However, ideal MHD implicitly overestimates the field strength in collapsing regions by neglecting reconnection and diffusion. 
Methods that include either ambipolar diffusion or Ohmic dissipation, but do not include radiative feedback, find that fragmentation may still occur in disks, particularly if the disk rotation rate is sufficiently high \citep{machida08, duffin09}. In sum, magnetic fields most likely reduce disk fragmentation, and  since we find no disk fragmentation when radiative feedback is included and both accretion and rotation rates are low, our conclusions would be unchanged.


The greatest uncertainty in our results arises from the absence of
protostellar outflows, which impact both the accretion and
luminosity. Observations of starless cores indicate that the core mass
function shape is similar to the stellar IMF but is
shifted to higher masses by a factor of three \citep{alves07,
  enoch08}. This difference is generally interpreted as an efficiency
factor reflecting the amount of gas launched and entrained in outflows
\citep{matzner00},
 although it does not imply a one-to-one correspondence between cores and stars. 

By making some simple assumptions we can estimate an
upper limit on the effect of neglecting outflows.
We first adopt a constant efficiency factor, $\epsilon = 1/3$,
and assume that the
accretion time remains constant (i.e., $M_*/\mdin$ = constant).
Under this transformation, a star in the
calculation with final mass $1\msun$ would instead accrete with $0.33
\mdin$ and have a final mass of
$0.33 \msun$. 
Since accretion dominates the luminosity for
most of the simulation (OKMK09), we approximate the total luminosity
by:
\begin{equation}
L_* \simeq L_{\rm acc} = {f_{\rm acc}{{G M_* \mdin }\over{R_*}}},
\end{equation}
where $f_{\rm acc}$ is the fraction of accretion energy that is radiated
away and $R_*$ is the protostellar radius. 
A power-law fit of a one-zone stellar evolution  model calibrated to
the evolutionary tracks of 
\citet{hosokawa09} gives $R_*
\propto M_*^{0.3}\mdin^{0.1}$ 
Altogether, the luminosity varies as $\epsilon^{1.6}$.
 
From equation 7, the disk temperature $T_{d} \propto (\epsilon^{-0.35}L_*/r_d^2)^{1/4}
 \propto \epsilon^{0.89}$, where $\avg{j_{\rm in}}$ is roughly independent
 of $\epsilon$ (e.g., \citealt{matzner00}) and the disk radius in a
 given core scales inversely with the mass: $r_d \propto \avg{j_{\rm in}}^2/M_*$.
 This gives $\xi \propto \epsilon^{-0.34}$ and
 $\Gamma \propto \epsilon$. 
Consequently, outflow mass loss shifts the
disks towards lower $\Gamma$ and higher $\xi$ values.
For $\epsilon = 1/3$, all but two RT systems stay to the left of the
 fragmentation line indicating that most systems remain stable.

If we relax the assumption that the accretion time is constant and
instead assume that $\mdin$ is constant (as in the case of a
collapsing isothermal sphere), then  $\xi \propto \epsilon^{-0.9}$ while
 $\Gamma$ is independent of $\epsilon$. Again only two previously stable
systems cross the fragmentation line; although, several additional
systems lie very close to the line. Reducing the core efficiency thus
has a mildly destabilizing affect on the disks.
These corrected values represent an upper
 limit on the change in the parameters, so it appears unlikely that
 including outflows in the simulations will decrease disk stability
 significantly and lead to increased multiplicity. 

We note that additional uncertainty may arise from the geometric aspect of
outflow cavites, which could affect the nature of the disk
illumination. This caveat is implicit in the use of equations (5)-(7).
It is also possible that the interaction of outflows with nearby filaments and
cores will in fact increase the amount of 
turbulent fragmentation, lower the characteristic protostellar mass
(\citealt{li10}; C. Hansen, 2010, private communication), 
and thus reduce the mean luminosity further.
Preliminary findings 
indicate that disk fragmentation remains uncommon in calculations 
where additional fragmentation is triggered by outflow interactions with turbulent
filaments (C. Hansen, 2010, private communication). Instead, the amount
of turbulent fragmentation is elevated relative
to disk fragmentation, which supports our thesis that disk
fragmentation around low-mass stars is rare. 

\subsection{Resolution Limitations}\label{sec-res}

In our calculation we take a conservative approach to the
representation of fragmentation with sink particles.
Although N-body gravitational interactions between particles may be modeled to
sub-grid cell accuracy, the gravitational interaction between the gas
and the stars is not well modeled when the separations are only a few
grid cells \citep{krumholz04}.
We choose to merge close particles rather than follow poorly resolved
interactions of the
particles and gas and thus introduce an  undetermined amount of error into
the gas and particle dynamics. As a result, we forfeit resolution of
binary systems with separations less than $\sim 200$ AU in the RT and
NRT simulations. In the RT case mergers happen relatively seldom and
the particles acheive at least a brown dwarf mass before
merging. These are all included in Figure \ref{dvst}, where their
initital pair separations suggest that the formation is real rather
than numerical.
It is also possible that unresolved fragmentation occurs close to the
primary star, although our RTC study suggests that this is unlikely in
most cases. 
 
Nonetheless, low
resolution gas fragments may form sink particles that would have become
thermally supported or dispersed at higher resolution. This situation
is more applicable to the NRT simulations, where demonstrably
unstable disks are produced when radiation feedback is absent. 
Although our sink particle method excludes formation in high-velocity
gas,
for the NRT run
it is likely that some cold fragmentation could be reduced by additional sink
criteria for the gravitational potential and gravitational boundedness
of clumps \citep{federrath10}.
The NRTC run does exhibit fragmentation at small scales
which may be indicative of an unresolved binary.
Thus, our
results are qualitatively supported by both higher resolution runs and
analysis of $\Gamma-\xi$ and $Q$. 

\subsection{Comparison with Observations and Previous Work}

These results affirm previous analytic and numerical work modeling
protostellar accretion disks. Our simulation sample is comprised of
low-mass stars with accretion rates below that required for disk
fragmentation as measured through $\xi$ and $\Gamma$.
The most massive RT star is $\sim$ 2.5
$\msun$, which is marginally in the regime where instability might be
expected according to \citet{kratter08}. However, they find that such a system would likely
not become unstable until 0.1-0.2 Myr or later. Lower mass stars are
expected to fragment at even later times or not at all, consistent with
the small amount of time that the RT stars spend to the right of the fragmentation line.
In contrast, the NRT and NRTC cases are highly discrepant
with the model predictions. It is clear that radiative
heating from the primary stars is a crucial component in simulations.

Most of the
core fragmentation occurs around separations of $\sim 0.01-0.02$ pc. Observations of
dense starless cores in Perseus suggest that the initital density
profile is relatively smooth outside the beam resolution of $\sim 1200$ AU
\citep{schnee10}.  
Two of the 11 cores in the sample are
found to have elongation on scales of a
few thousand AU that may be indicative of unresolved
fragmentation. 
A very smooth density distribution could preclude wide fragmentation as
a means for forming binary systems. 
However, our turbulence-produced fragments are rare and 
would be missed at the resolution of \citet{schnee10}. 
It is also possible that the observed starless cores may never form protostars or are younger than the objects we focus on in
this paper and may undergo fragmentation sometime in the future.

\citet{maury10} probe the multiplicity of Class 0 objects with
sensitivity down to 50
AU separations. Of the five sources in their sample, only one shows evidence of a
companion. This potential binary has a separation of $\sim$ 1900 AU, which suggests that
it did not result from disk fragmentation. Their findings, together with
\citet{looney00}, are consistent with our prediction that low-mass
protostars are not members of high-order multiple systems and that any companions
mostly likely originate from turbulent core fragmentation and form with a large initital separation.

Low-mass star systems like those modeled here are observed to have a
much lower binary fraction than higher mass stars \citep{lada06}. 
In particular, only 30\% of M dwarfs, the most common stellar type,
have lower mass companions.
However, both
simulations and analytic work predict that the more massive disks
around high-mass stars undergo fragmentation leading to additional
stellar companions \citep{kratter06,krumholz07a}. A
significant fraction of the difference may be due simply to the absence of
disk fragmentation around low mass stars. OKMK09 report a single star
fraction of $0.8 +0.2/-0.4$ or $0.5\pm 3$ assuming all mergers result
instead in a binary. These are likely upper and lower limits,
respectively, on the actual single star fraction, since the
distribution may evolve over time due to interactions between stars 
\citep{duchene07}.
Although the statistics are poor, these values are consistent with the
single star fraction of 0.7 \citep{lada06}. 

Using the simulations we can make predictions for the initial range of
binary separations. 
Excluding binaries
below the simulation resolution, these are in fact much larger
than the typical separation of $\sim 50$
AU, although there is a wide distribution around the mean, which varies
with stellar mass and possibly from region to region
\citep{duquennoy91,fisher04};
 \cite{raghavan10} find a 
gaussian distribution for periods peaked at $10^{5.03}$ days, implying $48$ AU separations for a total system mass of $1.5\msun$.
This suggests that
significant evolution of the separations must take place after the
    core fragments.  
The discrepancy is even larger for very low-mass binaries, whose
typical maximum separations are $1450 (M_{\rm tot}/\msun)^2$\,AU for systems
$\ltsim 0.6 \msun$ \citep{Burgasser03,Lafren08},
although wide binary exceptions do exist \citep{Radigan08, Radigan09}.   
We favor a scenario 
in which close low-mass binaries form via interaction with the ongoing
 infall 
 and between the two accretion disks; loosely bound companions
 can be stripped by close encounters within the protocluster.  
(The initial subvirial velocity dispersion of the stars predisposes
young clusters to dynamical interactions that will impact the final
binary distribution; \citealt{bate03, Offner09b}.)   


\section{Conclusions} \label{conclusions}

In this work, we revisit the turbulent, self-gravitating
radiation-hydrodynamic calculations performed by Offner et al.~(2009) in order to characterize the
protostellar systems in terms of the
thermal parameter, $\xi$, and the rotational parameter, $\Gamma$.  
We first use the high-resolution
simulations to confirm the stability
criteria derived in KMKK10 under non-turbulent, idealized conditions.
The high-resolution simulations further demonstrate that the dimensionless
parameters are converged at lower resolution and, thus, they  
are sufficient to describe the
evolutionary state of the protostellar disks even in the case where
the scale height and disk structure are not well resolved and where the
infall rate is fluctuating and turbulent. 

As expected, we find that $\xi$ and $\Gamma$ are distinct in the cases with and
without radiative feedback. In the former case, fragmentation occurs
preferentially on scales of $\sim$1000 AU rather than within disks.
These two
parameters indicate that protostellar accretion disks around low-mass
protostars will be stable against
gravitational fragmentation for nearly all times.
As an independent test, we perform a semi-analytic
analysis using the simulated accretion rates and luminosities, which
confirms that these disks are stable 
out to large radii.

Under a scenario of stable accretion disks, 
low mass 
 binary systems arise as a result of multiple
collapse events in a turbulent core. It follows that the observed multiplicity of
low-mass stellar systems arises predominatly due to turbulent
fragmentation in the parent core.

\acknowledgements{ We thank the anonymous referee for useful
  suggestions, which have improved the manuscript.
This research has been supported by the NSF through
  the grant AST-0901055 (SSRO). KMK is supported in part by an Ontario
  Graduate Scholarship. CDM is supported
by NSERC and an Ontario Early Researcher Award. RIK is supported by NASA through ATFP grant
  NNX09AK31G; the NSF through grant AST-0908553 and the US Department
  of Energy at the Lawrence Livermore National Laboratory under
  contract DE-AC52-07NA 27344. MRK acknowledges support from: an Alfred P.Sloan Fellowship; NASA through ATFP grant NNX09AK31G; NASA as part of the Spitzer Theoretical Research Program, through a contract issued by the JPL; the National Science Foundation through grant AST-0807739.}


\begin{thebibliography}{70}
\expandafter\ifx\csname natexlab\endcsname\relax\def\natexlab#1{#1}\fi

\bibitem[{{Adams} {et~al.}(1989){Adams}, {Ruden}, \& {Shu}}]{ARS89}
{Adams}, F.~C., {Ruden}, S.~P., \& {Shu}, F.~H. 1989, \apj, 347, 959

\bibitem[{{Alves} {et~al.}(2007){Alves}, {Lombardi}, \& {Lada}}]{alves07}
{Alves}, J., {Lombardi}, M., \& {Lada}, C.~J. 2007, \aap, 462, L17

\bibitem[{{Andre} \& {Montmerle}(1994)}]{andre94}
{Andre}, P. \& {Montmerle}, T. 1994, \apj, 420, 837

\bibitem[{{Bate}(2009)}]{bate09}
{Bate}, M.~R. 2009, \mnras, 392, 1363

\bibitem[{{Bate} {et~al.}(2003){Bate}, {Bonnell}, \& {Bromm}}]{bate03}
{Bate}, M.~R., {Bonnell}, I.~A., \& {Bromm}, V. 2003, \mnras, 339, 577

\bibitem[{{Bate} {et~al.}(2000){Bate}, {Bonnell}, {Clarke}, {Lubow}, {Ogilvie},
  {Pringle}, \& {Tout}}]{bate00}
{Bate}, M.~R., {Bonnell}, I.~A., {Clarke}, C.~J., {Lubow}, S.~H., {Ogilvie},
  G.~I., {Pringle}, J.~E., \& {Tout}, C.~A. 2000, \mnras, 317, 773

\bibitem[{{Boley} {et~al.}(2007){Boley}, {Durisen}, {Nordlund}, \&
  {Lord}}]{boley07}
{Boley}, A.~C., {Durisen}, R.~H., {Nordlund}, {\AA}., \& {Lord}, J. 2007, \apj,
  665, 1254

\bibitem[{{Bonnell} \& {Bate}(1994)}]{bonnell94}
{Bonnell}, I.~A. \& {Bate}, M.~R. 1994, \mnras, 269, L45

\bibitem[{{Burgasser} {et~al.}(2003){Burgasser}, {Kirkpatrick}, {Reid},
  {Brown}, {Miskey}, \& {Gizis}}]{Burgasser03}
{Burgasser}, A.~J., {Kirkpatrick}, J.~D., {Reid}, I.~N., {Brown}, M.~E.,
  {Miskey}, C.~L., \& {Gizis}, J.~E. 2003, \apj, 586, 512

\bibitem[{{Cai} {et~al.}(2008){Cai}, {Durisen}, {Boley}, {Pickett}, \&
  {Mej{\'{\i}}a}}]{cai08}
{Cai}, K., {Durisen}, R.~H., {Boley}, A.~C., {Pickett}, M.~K., \&
  {Mej{\'{\i}}a}, A.~C. 2008, \apj, 673, 1138

\bibitem[{{Clarke}(2009)}]{clarke09}
{Clarke}, C.~J. 2009, \mnras, 396, 1066

\bibitem[{{Duch{\^e}ne} {et~al.}(2007){Duch{\^e}ne}, {Delgado-Donate},
  {Haisch}, {Loinard}, \& {Rodr{\'{\i}}guez}}]{duchene07}
{Duch{\^e}ne}, G., {Delgado-Donate}, E., {Haisch}, Jr., K.~E., {Loinard}, L.,
  \& {Rodr{\'{\i}}guez}, L.~F. 2007, Protostars and Planets V, 379

\bibitem[{{Duffin} \& {Pudritz}(2009)}]{duffin09}
{Duffin}, D.~F. \& {Pudritz}, R.~E. 2009, \apjl, 706, L46

\bibitem[{{Duquennoy} \& {Mayor}(1991)}]{duquennoy91}
{Duquennoy}, A. \& {Mayor}, M. 1991, \aap, 248, 485

\bibitem[{{Enoch} {et~al.}(2008){Enoch}, {Evans}, {Sargent}, {Glenn},
  {Rosolowsky}, \& {Myers}}]{enoch08}
{Enoch}, M.~L., {Evans}, II, N.~J., {Sargent}, A.~I., {Glenn}, J.,
  {Rosolowsky}, E., \& {Myers}, P. 2008, \apj, 684, 1240

\bibitem[{{Evans} {et~al.}(2009){Evans}, {Dunham}, {J{\o}rgensen}, {Enoch},
  {Mer{\'{\i}}n}, {van Dishoeck}, {Alcal{\'a}}, {Myers}, {Stapelfeldt},
  {Huard}, {Allen}, {Harvey}, {van Kempen}, {Blake}, {Koerner}, {Mundy},
  {Padgett}, \& {Sargent}}]{evans09}
{Evans}, N.~J., {Dunham}, M.~M., {J{\o}rgensen}, J.~K., {Enoch}, M.~L.,
  {Mer{\'{\i}}n}, B., {van Dishoeck}, E.~F., {Alcal{\'a}}, J.~M., {Myers},
  P.~C., {Stapelfeldt}, K.~R., {Huard}, T.~L., {Allen}, L.~E., {Harvey}, P.~M.,
  {van Kempen}, T., {Blake}, G.~A., {Koerner}, D.~W., {Mundy}, L.~G.,
  {Padgett}, D.~L., \& {Sargent}, A.~I. 2009, \apjs, 181, 321

\bibitem[{{Federrath} {et~al.}(2010){Federrath}, {Banerjee}, {Clark}, \&
  {Klessen}}]{federrath10}
{Federrath}, C., {Banerjee}, R., {Clark}, P.~C., \& {Klessen}, R.~S. 2010,
  \apj, 713, 269

\bibitem[{{Fisher}(2004)}]{fisher04}
{Fisher}, R.~T. 2004, \apj, 600, 769

\bibitem[{{Foster} \& {Chevalier}(1993)}]{foster93}
{Foster}, P.~N. \& {Chevalier}, R.~A. 1993, \apj, 416, 303

\bibitem[{{Gammie}(2001)}]{gammie01}
{Gammie}, C.~F. 2001, \apj, 553, 174

\bibitem[{{Goldreich} \& {Lynden-Bell}(1965)}]{GLB1965}
{Goldreich}, P. \& {Lynden-Bell}, D. 1965, \mnras, 130, 125

\bibitem[{{Goodwin} {et~al.}(2007){Goodwin}, {Kroupa}, {Goodman}, \&
  {Burkert}}]{goodwin07}
{Goodwin}, S.~P., {Kroupa}, P., {Goodman}, A., \& {Burkert}, A. 2007,
  Protostars and Planets V, 133

\bibitem[{{Goodwin} {et~al.}(2004){Goodwin}, {Whitworth}, \&
  {Ward-Thompson}}]{goodwin04}
{Goodwin}, S.~P., {Whitworth}, A.~P., \& {Ward-Thompson}, D. 2004, \aap, 414,
  633

\bibitem[{{Hale}(1994)}]{hale94}
{Hale}, A. 1994, \aj, 107, 306

\bibitem[{{Hennebelle} \& {Teyssier}(2008)}]{hennebelle08}
{Hennebelle}, P. \& {Teyssier}, R. 2008, \aap, 477, 25

\bibitem[{{Hosokawa} \& {Omukai}(2009)}]{hosokawa09}
{Hosokawa}, T. \& {Omukai}, K. 2009, \apj, 691, 823

\bibitem[{{Howe} \& {Clarke}(2009)}]{howe09}
{Howe}, K.~S. \& {Clarke}, C.~J. 2009, \mnras, 392, 448

\bibitem[{{Jensen} {et~al.}(2004){Jensen}, {Mathieu}, {Donar}, \&
  {Dullighan}}]{jensen04}
{Jensen}, E.~L.~N., {Mathieu}, R.~D., {Donar}, A.~X., \& {Dullighan}, A. 2004,
  \apj, 600, 789

\bibitem[{{Klessen} \& {Burkert}(2000)}]{klessen00}
{Klessen}, R.~S. \& {Burkert}, A. 2000, \apjs, 128, 287

\bibitem[{{Kratter} \& {Matzner}(2006)}]{kratter06}
{Kratter}, K.~M. \& {Matzner}, C.~D. 2006, \mnras, 373, 1563

\bibitem[{{Kratter} {et~al.}(2008){Kratter}, {Matzner}, \&
  {Krumholz}}]{kratter08}
{Kratter}, K.~M., {Matzner}, C.~D., \& {Krumholz}, M.~R. 2008, \apj, 681, 375

\bibitem[{{Kratter} {et~al.}(2010{\natexlab{a}}){Kratter}, {Matzner},
  {Krumholz}, \& {Klein}}]{kratter10}
{Kratter}, K.~M., {Matzner}, C.~D., {Krumholz}, M.~R., \& {Klein}, R.~I.
  2010{\natexlab{a}}, \apj, 708, 1585

\bibitem[{{Kratter} {et~al.}(2010{\natexlab{b}}){Kratter}, {Murray-Clay}, \&
  {Youdin}}]{kratter10b}
{Kratter}, K.~M., {Murray-Clay}, R.~A., \& {Youdin}, A.~N. 2010{\natexlab{b}},
  \apj, 710, 1375

\bibitem[{{Krumholz} {et~al.}(2007){Krumholz}, {Klein}, \&
  {McKee}}]{krumholz07a}
{Krumholz}, M.~R., {Klein}, R.~I., \& {McKee}, C.~F. 2007, \apj, 656, 959

\bibitem[{{Krumholz} {et~al.}(2004){Krumholz}, {McKee}, \&
  {Klein}}]{krumholz04}
{Krumholz}, M.~R., {McKee}, C.~F., \& {Klein}, R.~I. 2004, \apj, 611, 399

\bibitem[{{Lada}(2006)}]{lada06}
{Lada}, C.~J. 2006, \apjl, 640, L63

\bibitem[{{Lafreni{\`e}re} {et~al.}(2008){Lafreni{\`e}re}, {Jayawardhana},
  {Brandeker}, {Ahmic}, \& {van Kerkwijk}}]{Lafren08}
{Lafreni{\`e}re}, D., {Jayawardhana}, R., {Brandeker}, A., {Ahmic}, M., \& {van
  Kerkwijk}, M.~H. 2008, \apj, 683, 844

\bibitem[{{Li} {et~al.}(2010){Li}, {Wang}, {Abel}, \& {Nakamura}}]{li10}
{Li}, Z., {Wang}, P., {Abel}, T., \& {Nakamura}, F. 2010, \apjl, 720, L26

\bibitem[{{Looney} {et~al.}(2000){Looney}, {Mundy}, \& {Welch}}]{looney00}
{Looney}, L.~W., {Mundy}, L.~G., \& {Welch}, W.~J. 2000, \apj, 529, 477

\bibitem[{{Lubow} \& {Ogilvie}(2000)}]{lubow00}
{Lubow}, S.~H. \& {Ogilvie}, G.~I. 2000, \apj, 538, 326

\bibitem[{{Machida} {et~al.}(2008){Machida}, {Tomisaka}, {Matsumoto}, \&
  {Inutsuka}}]{machida08}
{Machida}, M.~N., {Tomisaka}, K., {Matsumoto}, T., \& {Inutsuka}, S. 2008,
  \apj, 677, 327

\bibitem[{{Matzner} \& {Levin}(2005)}]{matzner05}
{Matzner}, C.~D. \& {Levin}, Y. 2005, \apj, 628, 817

\bibitem[{{Matzner} \& {McKee}(2000)}]{matzner00}
{Matzner}, C.~D. \& {McKee}, C.~F. 2000, \apj, 545, 364

\bibitem[{{Maury} {et~al.}(2010){Maury}, {Andr{\'e}}, {Hennebelle}, {Motte},
  {Stamatellos}, {Bate}, {Belloche}, {Duch{\^e}ne}, \& {Whitworth}}]{maury10}
{Maury}, A.~J., {Andr{\'e}}, P., {Hennebelle}, P., {Motte}, F., {Stamatellos},
  D., {Bate}, M., {Belloche}, A., {Duch{\^e}ne}, G., \& {Whitworth}, A. 2010,
  \aap, 512, A40+

\bibitem[{{McKee} \& {Tan}(2003)}]{mckee03}
{McKee}, C.~F. \& {Tan}, J.~C. 2003, \apj, 585, 850

\bibitem[{{Melo} {et~al.}(2001){Melo}, {Covino}, {Alcal{\'a}}, \&
  {Torres}}]{melo01}
{Melo}, C.~H.~F., {Covino}, E., {Alcal{\'a}}, J.~M., \& {Torres}, G. 2001,
  \aap, 378, 898

\bibitem[{{Mitchell}(1767)}]{Mitchell1767}
{Mitchell}, J. 1767, Philos. Trans. R. Soc., 76 , 97

\bibitem[{{Monin} {et~al.}(2006){Monin}, {M{\'e}nard}, \& {Peretto}}]{monin06}
{Monin}, J., {M{\'e}nard}, F., \& {Peretto}, N. 2006, \aap, 446, 201

\bibitem[{{Nelson}(2006)}]{nelson06}
{Nelson}, A.~F. 2006, \mnras, 373, 1039

\bibitem[{{Offner} {et~al.}(2009{\natexlab{a}}){Offner}, {Hansen}, \&
  {Krumholz}}]{Offner09b}
{Offner}, S.~S.~R., {Hansen}, C.~E., \& {Krumholz}, M.~R. 2009{\natexlab{a}},
  \apjl, 704, L124

\bibitem[{{Offner} {et~al.}(2008){Offner}, {Klein}, \& {McKee}}]{offner08}
{Offner}, S.~S.~R., {Klein}, R.~I., \& {McKee}, C.~F. 2008, \apj, 686, 1174

\bibitem[{{Offner} {et~al.}(2009{\natexlab{b}}){Offner}, {Klein}, {McKee}, \&
  {Krumholz}}]{offner09}
{Offner}, S.~S.~R., {Klein}, R.~I., {McKee}, C.~F., \& {Krumholz}, M.~R.
  2009{\natexlab{b}}, \apj, 703, 131

\bibitem[{{Price} \& {Bate}(2007)}]{price07}
{Price}, D.~J. \& {Bate}, M.~R. 2007, \mnras, 377, 77

\bibitem[{{Radigan} {et~al.}(2008){Radigan}, {Lafreni{\`e}re}, {Jayawardhana},
  \& {Doyon}}]{Radigan08}
{Radigan}, J., {Lafreni{\`e}re}, D., {Jayawardhana}, R., \& {Doyon}, R. 2008,
  \apj, 689, 471

\bibitem[{{Radigan} {et~al.}(2009){Radigan}, {Lafreni{\`e}re}, {Jayawardhana},
  \& {Doyon}}]{Radigan09}
---. 2009, \apj, 698, 405

\bibitem[{{Rafikov}(2005)}]{rafikov05}
{Rafikov}, R.~R. 2005, \apjl, 621, L69

\bibitem[{{Schnee} {et~al.}(2010){Schnee}, {Enoch}, {Johnstone}, {Culverhouse},
  {Leitch}, {Marrone}, \& {Sargent}}]{schnee10}
{Schnee}, S., {Enoch}, M., {Johnstone}, D., {Culverhouse}, T., {Leitch}, E.,
  {Marrone}, D., \& {Sargent}, A. 2010, Accepted to ApJ

\bibitem[{{Scholz} {et~al.}(2010){Scholz}, {Wood}, {Wilner}, {Jayawardhana},
  {Delorme}, {Garatti}, {Ivanov}, {Saviane}, \& {Whitney}}]{scholz10}
{Scholz}, A., {Wood}, K., {Wilner}, D., {Jayawardhana}, R., {Delorme}, P.,
  {Garatti}, A.~C.~o., {Ivanov}, V.~D., {Saviane}, I., \& {Whitney}, B. 2010,
  ArXiv e-prints

\bibitem[{{Sellwood} \& {Carlberg}(1984)}]{sellwood84}
{Sellwood}, J.~A. \& {Carlberg}, R.~G. 1984, \apj, 282, 61

\bibitem[{{Shakura} \& {Sunyaev}(1973)}]{SS1973}
{Shakura}, N.~I. \& {Sunyaev}, R.~A. 1973, \aap, 24, 337

\bibitem[{{Shu}(1977)}]{shu77}
{Shu}, F.~H. 1977, \apj, 214, 488

\bibitem[{{Shu} {et~al.}(2004){Shu}, {Li}, \& {Allen}}]{shu04}
{Shu}, F.~H., {Li}, Z., \& {Allen}, A. 2004, \apj, 601, 930

\bibitem[{{Shu} {et~al.}(1990){Shu}, {Tremaine}, {Adams}, \& {Ruden}}]{shu1990}
{Shu}, F.~H., {Tremaine}, S., {Adams}, F.~C., \& {Ruden}, S.~P. 1990, \apj,
  358, 495

\bibitem[{{Stamatellos} \& {Whitworth}(2009)}]{stamatellos09}
{Stamatellos}, D. \& {Whitworth}, A.~P. 2009, \mnras, 392, 413

\bibitem[{{Stone} {et~al.}(1998){Stone}, {Ostriker}, \& {Gammie}}]{stone98}
{Stone}, J.~M., {Ostriker}, E.~C., \& {Gammie}, C.~F. 1998, \apjl, 508, L99

\bibitem[{{Tohline}(2002)}]{tohline02}
{Tohline}, J.~E. 2002, \araa, 40, 349

\bibitem[{{Toomre}(1964)}]{toomre64}
{Toomre}, A. 1964, \apj, 139, 1217

\bibitem[{{Truelove} {et~al.}(1997){Truelove}, {Klein}, {McKee}, {Holliman},
  {Howell}, \& {Greenough}}]{truelove97}
{Truelove}, J.~K., {Klein}, R.~I., {McKee}, C.~F., {Holliman}, II, J.~H.,
  {Howell}, L.~H., \& {Greenough}, J.~A. 1997, \apjl, 489, L179+

\bibitem[{{Walch} {et~al.}(2009){Walch}, {Burkert}, {Whitworth}, {Naab}, \&
  {Gritschneder}}]{walch09}
{Walch}, S., {Burkert}, A., {Whitworth}, A., {Naab}, T., \& {Gritschneder}, M.
  2009, \mnras, 400, 13

\bibitem[{{White} \& {Hillenbrand}(2004)}]{white04}
{White}, R.~J. \& {Hillenbrand}, L.~A. 2004, \apj, 616, 998

\end{thebibliography}

\end{document}